\documentstyle[12pt,epsf,epsfig,wrapfig]{article}
\newcommand{\ba}{\begin{eqnarray}}
\newcommand{\ea}{\end{eqnarray}}

\newcommand{\beq}{\begin{equation}}
\newcommand{\eeq}{\end{equation}}
\newcommand{\beqs}{\begin{eqnarray}}
\newcommand{\eeqs}{\end{eqnarray}}

\begin{document}

\begin{center}
{\bfseries Generalized Parton Distributions and Nucleon Form Factors }

\vskip 5mm
\underline{O.V. Selyugin}$^{1 \dag}$and
O.V. Teryaev$^{1 \dag \dag }$

\vskip 5mm
{\small
(1) {\it
Bogoliubov Laboratory of Theoretical Physics, \\
Joint Institute for Nuclear Research,
141980 Dubna, Moscow region, Russia
}\\
$\dag$ {\it
E-mail: selugin@theor.jinr.ru
} \\
$\dag \dag$ {\it
E-mail: teryaev@theor.jinr.ru
}

}
\end{center}

\vskip 5mm
\begin{abstract}
   The Dirac and Pauli form factors of the proton and neutron
   are obtained in the framework of the generalized parton distributions (GPDs)
    with some simple momentum transfer dependence.
    It is shown that both sets of the existing experimental data on the form factors,
   obtained by the Rosenbluth and polarization transfer, can be  described
    by changing only the slope of the GPDs $E$. The description of neutron form factors
    is substantially better when the proton data obtained by the studies of polarization transfer are used.
\end{abstract}

       \section{Introduction}

   The determination of the hadron structure is related with our understanding
   of the non-perturbative properties of the QCD.
   Generalized parton distributions (GPDs) \cite{GPDs}
   for $\xi =0 $ provide information about the distribution of the partons
   in impact parameter space \cite{Burk1}.
   It is correlated with $t$-dependence of  GPDs.
   Now we cannot obtain this dependence from the first principles;
   instead, it may be obtained from the phenomenological description
   of the nucleon electromagnetic form-factors.

  Following \cite{R98},  we limit ourselves to the case of GPDs with $\xi=0$
  corresponding to the non-forward parton densities so that
the form factors can be represented as
\ba
 F_{1}^q (t) = \int^{1}_{0} \ dx  \ {\cal{ H}}^{q} (x, t), \ \ \
 F_{2}^q (t) = \int^{1}_{0} \ dx \  {\cal{E}}^{q} (x, t),
\ea
  We assume the validity of Gaussian ansatz  which was used in  \cite{R98} to describe the form factors of proton.
However, this ansatz leads to a faster decrease in  $F_1$ at larger momentum transfer.
Although this region is, strictly speaking, outside the domain of validity of QCD factorization involving GPDs,
one may consider also the problem of $t$-dependence of  GPDs at large $t$ \cite{Burk04}.
It was shown that at  large $x  \rightarrow 1$ and momentum transfer
  the behavior of GPDs
  requires a (larger) power dependence on $(1-x)$ in the $t-$ dependent exponent:
\ba
{\cal{H}}^{q} (x,t) \  \sim  exp [ a \ (1-x)^n \ t ] \ q(x).
\ea
with $n \geq 2$. It was noted that $n=2$ naturally gives rise to Drell-Yan-West duality between parton distributions
at large $x$ and the form factors.
Various more elaborated parameterizations were considered later, see e.g.  \cite{Kroll04}.

 \section{Momentum transfer dependence of GPDs and proton form factors }

 Our proposal consists in the attempt to find a simple ansatz which will be good enough
 to describe the form factors of the proton and neutron taking into account
  a number of  new data that have appeared in the last years.
  Let us keep the simple Gaussian ansatz but using some new conditions.
To support the proposal \cite{R98} and \cite{Burk04}
we chose the t-dependence of GPDs in the form
 \ba
  {\cal{H}}^{q} (x,t) \  = q(x) \   exp [ a_{+}  \  \frac{(1-x)^2}{x^{b} } \ t ]; \ \  \
  {\cal{E}}^{q} (x,t) \  =  {\cal{E}}^{q} (x) \
      exp [ a_{-}  \  \frac{(1-x)^2}{x^{b}}  \ t ].
  \ea
 with the free parameters $b=0.4$ (determined mostly by the power $2$ of the factor $1-x$),
 $a_{\pm}$ ($a_{+} $ - for ${\cal{H}}$ and $a_{-} $ - for ${\cal{E}}$). All these
 parameters were fixed by analyzing the data on the ratio of proton Pauli and Dirac form-factors.
 The function $q(x)$ was taken in the same normalization point $\mu^2 = 1 \ $GeV$^2$ as in \cite{R04},
  which is based on the MRST2002 global fit \cite{MRST02}.
 In all our calculations we restricted ourselves
 to the contributions of $u$ and $d$ quarks
  in  ${\cal{H}}^{q}$ and ${\cal{E}}^{q}$
   with
 $   {\cal{E}}^{u} (x) \  = k_{u}/N_{u} (1-x)^{\kappa_1} \ u(x), \ \ \
 {\cal{E}}^{d} (x) \  = \frac{k_d}{N_d} (1-x)^{\kappa_2} \ d(x)$,
   (where $\kappa_1 =1.53$ and $\kappa_2=0.31$ \cite{R04})
   According to  the normalization of the Sachs form factors,
    we have
   $k_u=1.673  , \ \ \  k_d=2.033,   \  \  \  \   \ N_u=1.53  , \ \ \  N_d=0.946   $.
   The parameters $a_{+} = 0.675$ and $a_{-} $ correspond to the
    two experimental methods of the determination of the ratio of the Pauli and Dirac form factors.
    Below we consider  version (I - polarization transfer method) leading to $a_{-} = 0.59$
    and version (II - Rosenbluth separation) leading to $a_{-} = 0.7$ .

    \begin{figure}[b!]
    \vspace{-1.5cm}
\begin{center}
\begin{tabular}{cc}
\mbox{\epsfig{figure=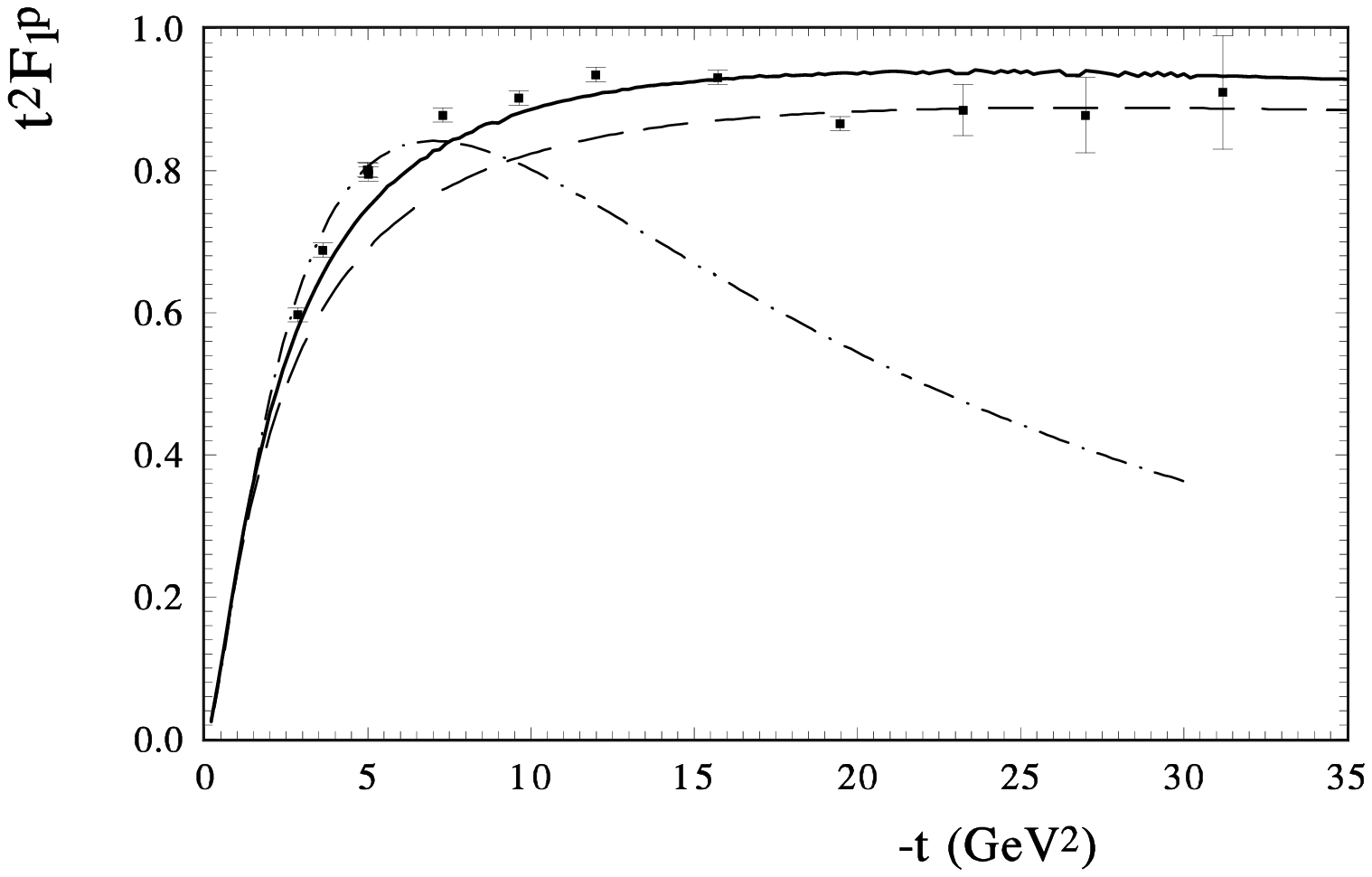,width=7cm,height=5cm}}& 
\mbox{\epsfig{figure=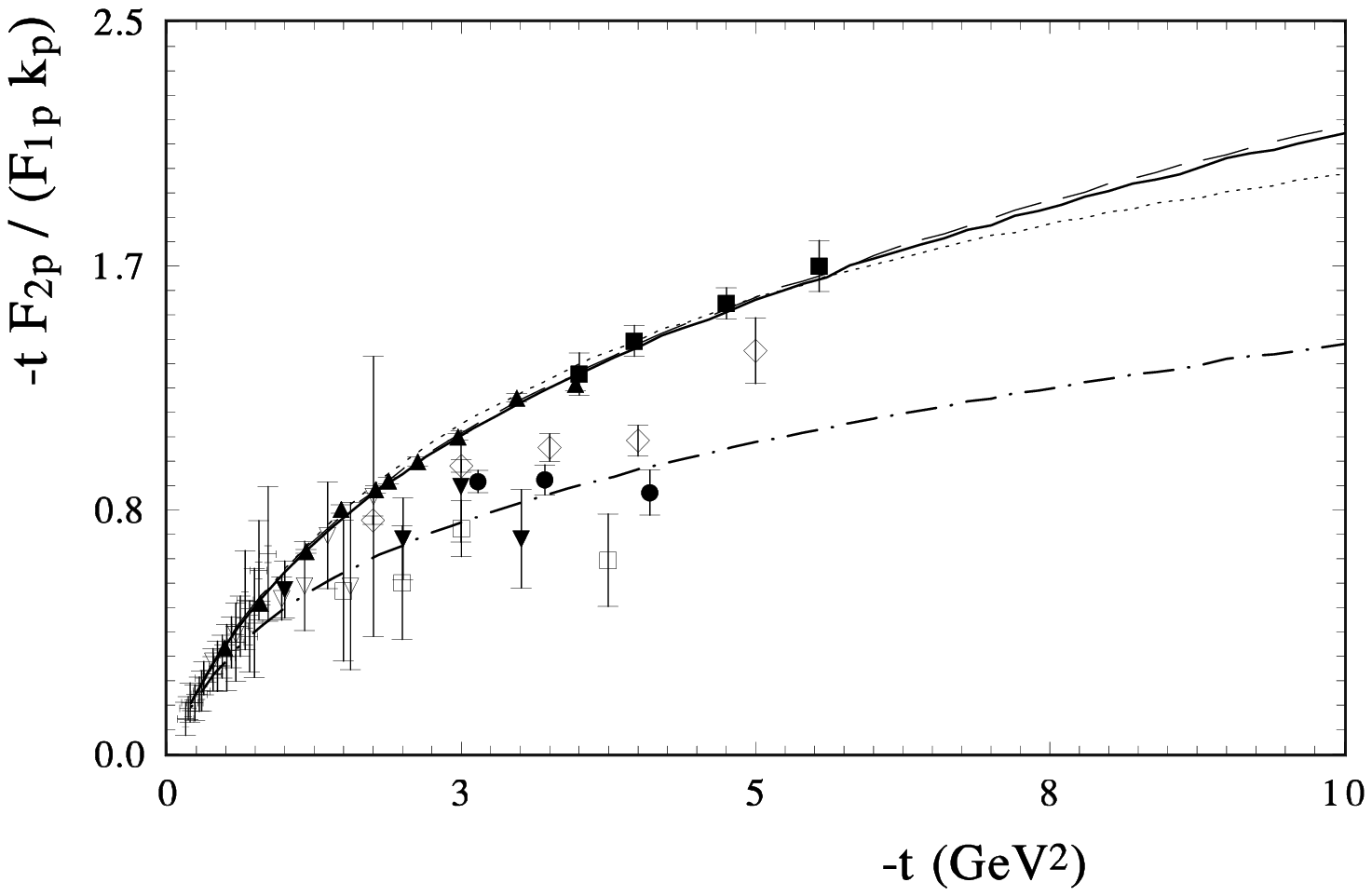,width=7cm,height=5cm}} \\
{\small{ {\bf(a)} }}&{\small{ {\bf(b)}}} 
\end{tabular} 
 \end{center}
{\small{\bf Figure 1a.}
Proton Dirac form factor multiplied by $t^2$
(hard line - the present work,   dot-dashed line - \cite{R98}; long-dashed line - \cite{R04};
  the  data for $F_1^{p}$ are from \cite{Sill93}.
}\\
{\small{\bf Figure 1b.} Ratio of the Pauli to  Dirac  proton form factors multiplied by $t$
(hard and dot-dashed lines correspond to version (I) and (II)) of the present  work ,   dotted line - \cite{Brodsky03}; long-dashed line - \cite{R04})
 ; the data  are from \cite{Jones00}.
   }
\end{figure}

The  proton Dirac form factor,  calculated in this work and multiplied by $t^2$,
is shown in Fig.1a in comparison with the other works (\cite{R04},\cite{Stol01}) and experimental data.
 One can see, that our calculations sufficiently well reproduces the behavior
 of experimental data not only at high $t$ but also at  low $t$.

 The ratio of the Pauli to the Dirac  proton form factors multiplied by $t$
 is shown in Fig.1b.
 There are  two different sets of experimental data. Firstly, one may extract the form factors of the proton
 from the unpolarised differential cross section by the Rosenbluth method.
 The other method uses the polarized differential
  cross section to obtain these form factors.
  In our model we can obtain the results
  of both methods by changing the slope of ${\cal{E}} $.
  So we examined two versions
  differing  by the slopes $a_{-}$.

  One can  now use the information on the neutron form factors
  in order to choose the
  more realistic version.

    \begin{figure}[b!]
\begin{center}
\vspace{-1.5cm}
\begin{tabular}{cc}
\mbox{\epsfig{figure=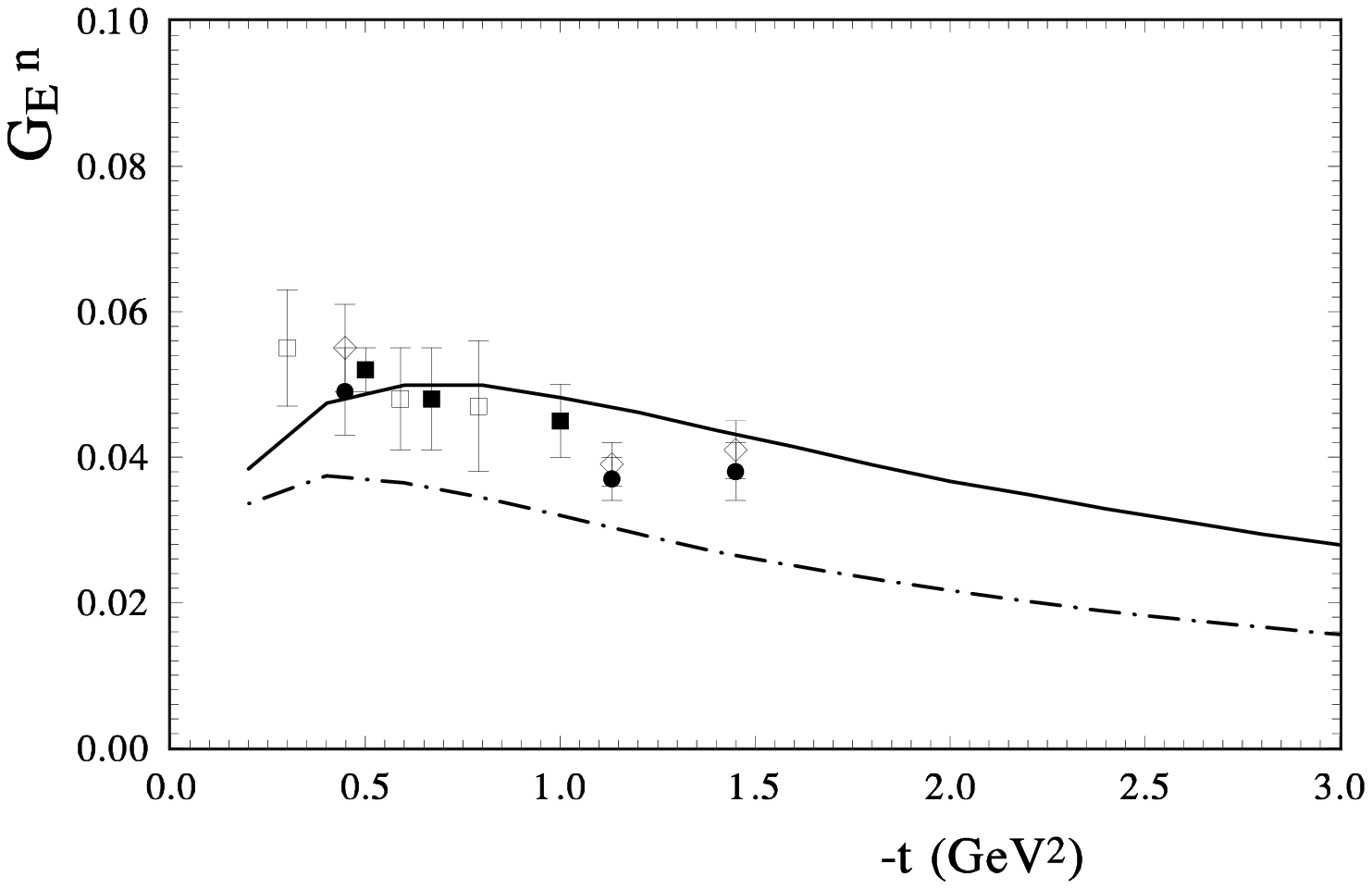,width=7cm,height=5cm}}& 
\mbox{\epsfig{figure=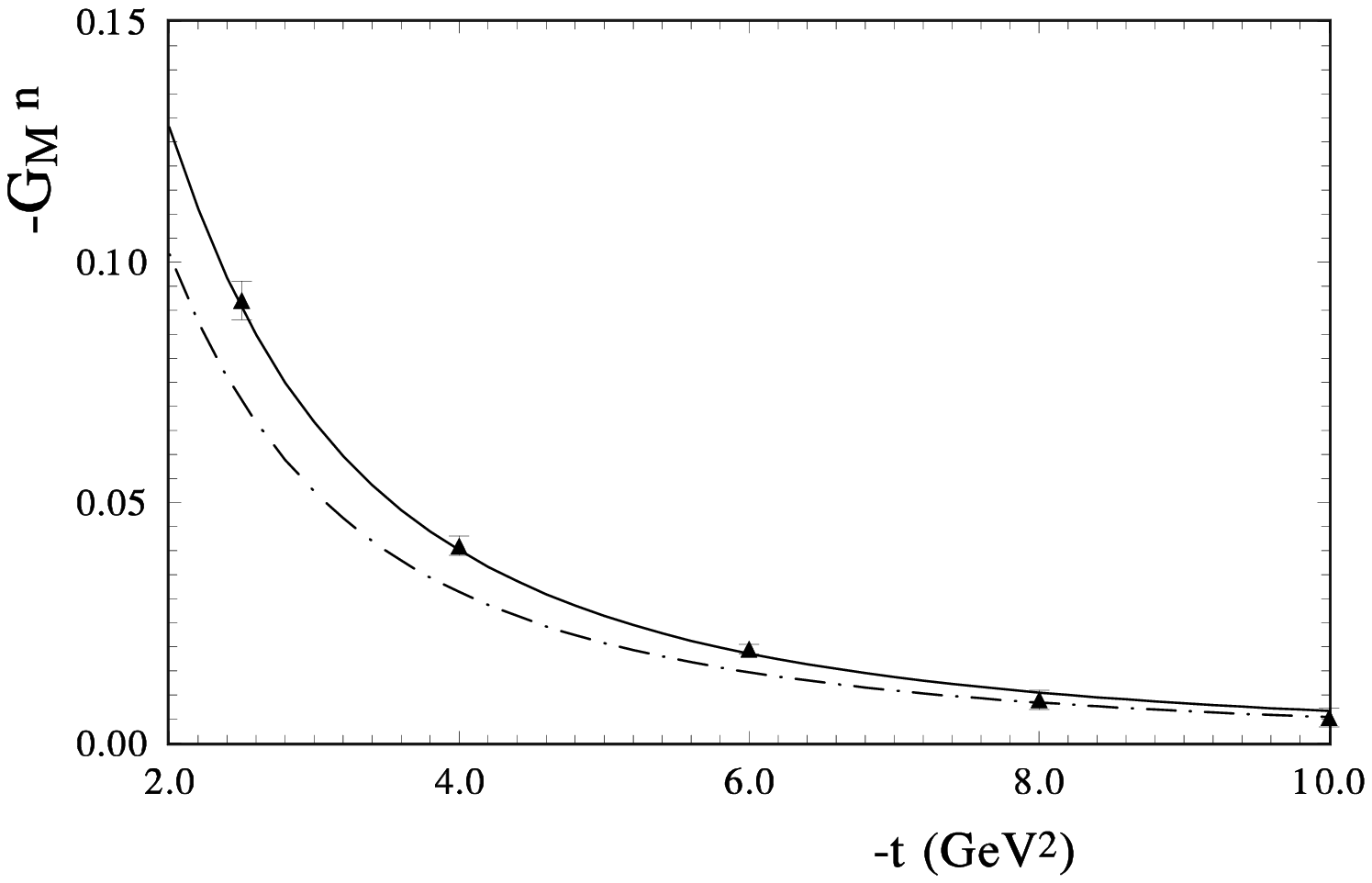,width=7cm,height=5cm}} \\
{\small{\bf(a)}}&{\small {\bf(b)}} 
\end{tabular} 
 \end{center}
{\small{\bf Figure 2a.}
$G_{E}^{n}$ (hard and dot-dashed lines correspond to version (I) and (II)); experimental data
from \cite{Plaster05}.
}\\
{\small{\bf Figure 2b.}
$G_{M}^{n}$ (hard and dot-dashed lines correspond to version (I) and (II)); experimental data
from \cite{Rock82}.
   }
\end{figure}


\section{Neutron form factors }

  Using the model  developed for proton we can calculate the neutron form factors.
  For that  the isotopic invariance can be used to relate the proton GPDs to the neutron ones,
  Hence, we do not change any parameters
 and preserve the same $t$-dependence of GPDs as in the case of  proton.

   Again, we take two values of the slope $a_{-} $
   as in the case of the proton form factors with the same size,
  which correspond to version $(I)$ and version $(II)$ below.

  Our  calculation of the $G_{E}^{n}$  is shown in Fig. 2a.

 Evidently, the first version is in better agreement
 with  experimental data. Therefore, neutron data support the results obtained by
 polarization transfer method.

%

This conclusion is supported by the  calculations of $G_{M}^{n}$ shown in Fig.2b.
  In this case,
   it is clearly seen that our parameterization normalized using the {\it proton} form-factors ratio from the polarization experiments
   describes these {\it neutron} data quite well.


\section{Conclusions }

    The proposed version of Gaussian $t$-dependence of GPDs  reproduces the electromagnetic structure of  the proton
    and neutron sufficiently well.
    We show that changing only the slope parameters $a_{-}$ of ${\cal{E}}^{q}$
    it is possible to obtain both the Rosenbluth and Polarization data on the ratio
    of Pauli and dirac electromagnetic proton form-factors. The description of neutron form-factors
    is essentially better with the slope parameter fitted to proton polarization transfer data.
    This is in accordance with the recent theoretical analysis \cite{Bystritskiy:2007hw}.

\vspace{0.5cm}

{\bf Acknowledgments}
We are indebted to M. Anselmino, S.V. Goloskokov, P. Kroll, E.~A.~Kuraev, E. Predazzi and E.~Tomasi-Gustafsson for useful discussions.
This work was partially supported by the Deutsche Forschungsgemeinschaft, grant 436 RUS 113/881/0, the Russian Foundation for Basic Research
(Grant 03-02-16816) and the Russian Federation Ministry of Education and Science (Grant MIREA 2.2.2.2.6546).


\end{document}